\documentclass[a4paper]{VisionStyle}
\usepackage{graphicx}

\begin{document}

\title{XMM-Newton observations of variability in the X-ray binaries of M31: flares, dips and a burst}

\author{R.\,Barnard\inst{1} \and U.\,Kolb\inst{1} \and 
  J. P.\,Osborne\inst{2} } 

\institute{
   Department of Physics \& Astronomy, The Open University, Walton Hall, Milton Keynes, England
\and 
  Department of Physics \& Astronomy, The University of Leicester, Leicester, England}

\maketitle 

\begin{abstract}

The {\em XMM-Newton} satellite has made three observations of the core of \object{M31} as part of a survey that will span the entire galaxy. The majority of the X-ray emission is contributed by point sources, chiefly X-ray Binaries (XB). Exciting early results from a survey of time variability in three XMM exposures of the M31 core reveal XB exhibiting  Sco\thinspace X-1-like flaring,   dips and a burst. The flaring source in M31 also exhibits variability in intensity and spectral hardness that is characteristic of a Z-source.     

\end{abstract}

\section{Introduction}

 At 760 kpc (\cite{rbarnard-C1:van00}) the Andromeda Galaxy (\object{M31}) is the nearest spiral galaxy to our own. The X-ray emission of M31 is dominated by point sources, either X-ray binaries or  supernova remnants with luminosities $\geq 10^{35}$ erg $s^{-1}$.  
X-ray binaries (XB) are close binary star systems in which one star is normal and the other is a neutron star or black hole. They are classified by the mass of the normal star--- high mass X-ray binaries (HMXB) are generally powered by wind accretion while low mass X-ray binaries (LMXB) are powered by Roche lobe overflow via disc accretion. HMXB occur in regions of recent star formation ($<10^{7}$ yr) while LMXB have longer evolutionary time scales and are associated with older stellar populations. There is little star formation in M31 and the core is dominated by old stars.

 Three {\em XMM-Newton} observations have been made of the core of M31, as part of a survey of the whole galaxy.  Only one X-ray source has in the core been identified with a SNR--- \object{X\thinspace 004327.9+411835} (\cite{rbarnard-C1:bla81}); hence the majority of X-ray sources will be XB. These observations yield information on  XB with luminosities $>$ a few 10$^{36}$ erg s$^{-1}$). X-ray binaries exhibit several types of variability on short  time-scales (i.e. milliseconds to hours), e.g. X-ray bursts, X-ray flares or dips as reviewed below. Hence a survey of the X-ray sources in M31 has been undertaken, in search of such variability in the combined  lightcurves of the EPIC MOS and PN detectors on board $XMM-Newton$. Early results are presented in this work. Variability of X-ray sources over 6 month time-scales is discussed by \cite*{rbarnard-C1:osb01} who also discovered a short period supersoft transient.


\section{Hard X-ray Flares}

Hard X-ray flares have been observed in only three LMXB in our galaxy---\object{ Sco\thinspace X-1}, \object{Sco\thinspace X-2} and \object{X\thinspace 1624--490} (\cite{rbarnard-C1:whi85}; \cite{rbarnard-C1:jw89}). Sco\thinspace X-1 and Sco\thinspace X-2 are classified as Z-sources, characterised by luminosities of $>10^{38}$ erg s$^{-1}$, a low inclination angle  and strongly correlated temporal and spectral variability; \object{Sco\thinspace X-1} and \object{Sco\thinspace X-2} exhibit active states when flaring occurs and quiescent states, known as the flaring branch and normal branch respectively. \object{X\thinspace  1624--490}  exhibits flaring and non-flaring states and a high luminosity, but it also exhibits periodic intensity dips due its high inclination angle  and so is classified as  a dipper.

Flaring involves an increase in the intensity and hardness over a period of a few hundred to a few thousand seconds before returning to the original spectral hardness and intensity. It seems likely that such flaring behaviour is characteristic of a Z-source.
\cite*{rbarnard-C1:hv87} classified six Galactic LMXB as Z-sources, characterised by high luminosity, low inclination angle and a 6 Hz quasi-periodic oscillation on the normal branch.  However, not all Z-sources flare on the flaring branch; Z-sources may be classified as \object{Sco\thinspace X-1} like (which flare) and \object{Cyg\thinspace X-2} like (which do not) (\cite{rbarnard-C1:hv87}).

To date \object{LMC\thinspace X-2} is the only known extra-galactic  Z-source (\cite{rbarnard-C1:sk00}).

\section{X-ray Bursts}

X-ray bursts are caused by thermonuclear burning of accreted material on the neutron star surface (e.g. \cite{rbarnard-C1:lam00}). The luminosity rapidly reaches the Eddington limit then decays exponentially; typical bursts have rise times of a few seconds and decay over tens or  hundreds of seconds; however, bursts have been observed with durations of several hours (e.g. \cite{rbarnard-C1:cor00}).
 The frequency of bursts is inversely proportional to the non-burst luminosity of the LMXB; X-ray bursting halts if the luminosity of the LMXB is normally greater than half the Eddington luminosity. No X-ray bursts have been detected outside our galaxy prior to these observations.

\section{ Dipping}

Accretion onto the neutron star in LMXB occurs via an accretion disc; the stream of matter flowing from the normal star collides with the outer edge of the accretion disc, causing a bulge to be created (\cite{rbarnard-C1:kra59}). In systems with a high inclination angle, the X-ray source may be periodically obscured by material in the bulge, causing photo-electric absorption, preferentially removing photons at lower energies (e.g. \cite{rbarnard-C1:ws82}); this is dipping.
No dips have been previously detected in LMXB outside our Galaxy.

\section{ The Observations}

Three observations of the core of \object{M31} were made with {\em XMM-Newton}:  June 26 2000 (Obs. 1),  December 27 2000 (Obs. 2) and  June 29 2001 (Obs. 3). The observations lasted 38 ks, 13 ks and 56 ks respectively.

 Lightcurves for the 2000, December observation  were extracted for all three EPIC instruments from each of the 35 brightest point sources in a broad energy band (0.3--10 keV) and narrower bands to highlight dips (0.3--2.5 keV), bursts (2.5--10 keV) and flares (4-10 keV). The extraction regions were circular with a 20'' radius; the equivalent background lightcurves were extracted for each source from nearby regions of the same size but free of point sources. 

The lightcurves from MOS1, MOS2 and the PN were combined for each source, and variability was sought with 100s, 200s and 400s time bins by fitting a constant intensity to the lightcurves.
 Four point sources exhibited striking variability, and lightcurves from the other two observations were obtained for them. They will be discussed in turn. None of these sources are associated with the globular clusters identified in M31  by \cite*{rbarnard-C1:bat87}.

\section{ A new flaring Z-source? }

\begin{figure}[!t]
\centering
\includegraphics[scale=0.3,angle=270]{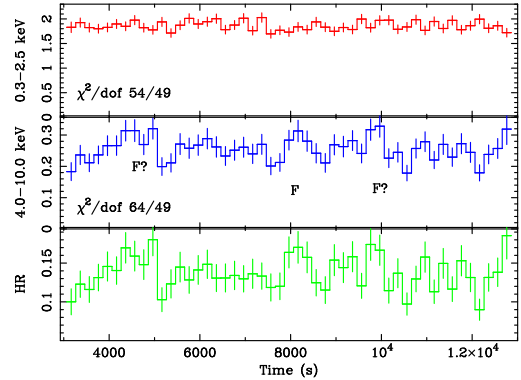}
\caption{$XMM$ X-ray lightcurves of  S1 from Obs. 1  in the 0.3--2.5 and 4--10 keV energy bands are shown in the top figure, along with the hardness ratio (i.e. the hard X-ray intensity divided by the soft X-ray intensity); flaring intervals are labelled F}\label{rbarnard-C1_16f}
\end{figure}

\begin{figure}[!t]
\centering
\includegraphics[scale=0.3,angle=270]{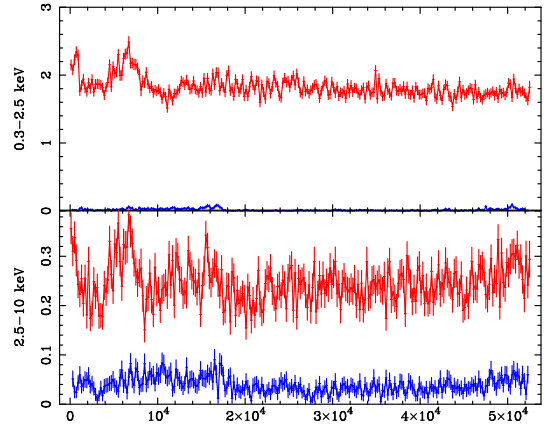}
\caption{ $XMM$ lightcurves of  S1 from  Obs. 3 exhibiting characteristic Z-source variability; source and background lightcurves (red and blue respectively) are shown for the low and high energy bands. Both source lightcurves show an increase in the intensity, which later returns to its original value; the amplitude of the enhancement is higher in the high energy lightcurve, meaning that the hardness is correlated with the intensity.}\label{rbarnard-C1_scof}
\end{figure}

The point source \object{X\thinspace 004238.6 + 411604} (hereafter known as S1) is the brightest source in the field of view. Possible hard flaring is exhibited in the first and second observations, and  striking variability in all energy bands  is exhibited  in the third observation, which resembles the flaring branch movement of Galactic Z-sources (\cite{rbarnard-C1:whi85}).  With 500 s time binning, the probability that the 2.5--10 keV variability during the flaring in observation 1 is significant  is $>$99.5\%. The 4--10 keV flaring in the first $XMM$ observation of S1 is shown  in Fig.~\ref{rbarnard-C1_16f}, while the likely flaring branch movement in the third observation of S1 is presented in  Fig.~\ref{rbarnard-C1_scof}.

\cite*{rbarnard-C1:tru02} report an absorbed flux for S1 of $\sim$4 $ \times 10^{-12}$ erg cm$^{-2}$ s$^{-1}$ in the 0.3--7 keV band. This corresponds to  a luminosity of $\sim$3 $ \times 10^{38}$ erg s$^{-1}$, consistent with the identification of S1 as a Z-source.

\section{ A candidate bursting dipper}

\begin{figure}[!t]
\centering
\includegraphics[scale=0.3, angle=270]{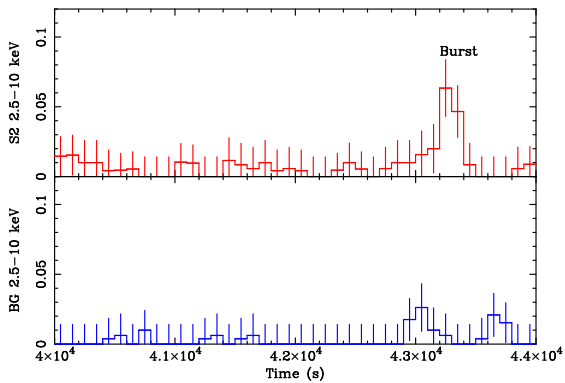}
\caption{$XMM$ lightcurves of  S2 (top panel) and a nearby background region (bottom panel) from Obs. 3; the burst occurred $\sim$43 ks after the start of the observation. The lightcurves are binned to 100 s, meaning that the peak of the burst is likely to be intrinsically much higher; the duration of the burst is $\sim$200s, which is typical}\label{rbarnard-C1_bst}
\end{figure}

A possible  X-ray burst has been identified in the lightcurve of  \object{XB\thinspace 004218.5+411223} (S2), and if confirmed would be  the first to be detected in an extra-galactic LMXB. It was detected in the 2.5--10 keV energy band, and occurred $\sim$43 ks into the 2001, June observation (Fig.~\ref{rbarnard-C1_bst}). The 2.5--10 keV lightcurve of S2 is compared to the 2.5--10 keV lightcurve of a nearby background region, to illustrate that the event is unique to S2.  The lightcurve is binned to 100 s, and so the peak intensity is probably  much diluted; however, the peak intensity is still $\sim$4 times the quiescent intensity. \cite*{rbarnard-C1:tru02} modelled the spectrum of S2 with an absorbed power law, obtaining an absorbed  flux of $\sim$3$ \times 10^{-13}$ erg cm$^{-2}$ s$^{-1}$ in the 0.3--7 keV band, corresponding to a luminosity of $\sim$ 2 $\times 10^{37}$ erg s$^{-1}$, which is within the luminosity bounds for a burster ($\sim$1--50\% of the  Eddington luminosity).

\begin{figure}[!t]
\centering
\includegraphics[scale=0.3,angle=270]{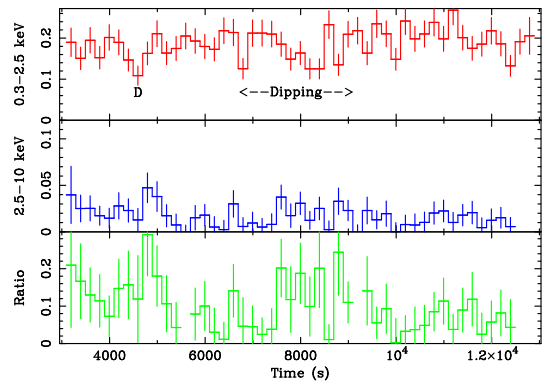}
\caption{Low and high energy Obs. 2 lightcurves of S2,  plus the ratio of their intensities. One possible dipping interval is explicitly labelled, another is labelled with D. The hardness ratio (here the 2.5--10 keV intensity divided by the 0.3--2.5 keV intensity) tends to be  higher in the dipping intervals than in the non-dip intervals, broadly consistent with photo-electric absorption}\label{rbarnard-C1_s2d}
\end{figure}

S2 is also a strong candidate to be classified as a dipping source; the low and high energy lightcurves are shown for the second observation of S2 in Fig.~\ref{rbarnard-C1_s2d}. The dipping is detected as a decrease in the low energy band intensity with a lesser variation in the high energy band, resulting in an increase in hardness; during dipping, the ratio of intensities of the 2.5--10 keV band to the 0.3--2.5 keV band doubles. There is no evidence of periodicity in this observation, indicating that the orbital period is longer than the duration of this short observation. The dipping is detected at a probability of $>$99.5\% as simple variability in the 0.3--2.5 keV lightcurve; however, the structure in the lightcurve increases the significance.

\section{ Possible soft flaring in a supersoft source }

\begin{figure}[!t]

\includegraphics[scale=0.3,angle=270]{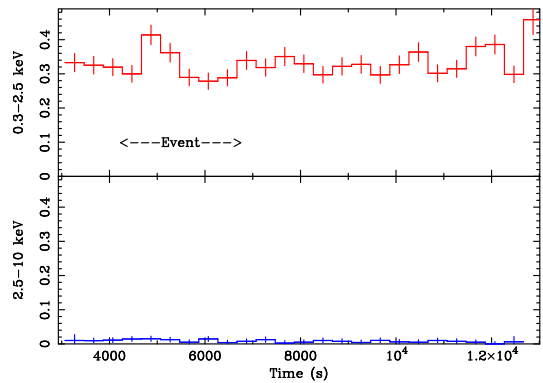}
\caption{Lightcurves of S3 from Obs 3.  in the 0.3--2.5 and 2.5--10 keV energy bands binned to 400 s, underlining the softness of the source. The soft intensity enhancement is labelled Event;  however, the intensity of the last time bin is also high. This soft excess may be due to some new kind of soft flaring; later spectral analysis should clarify the matter}\label{rbarnard-C1_ev}
\end{figure}

 \object{X\thinspace 004253.1+411530} (S3) shows long-term intensity variations: \cite*{rbarnard-C1:sup97} report that its luminosity in a ROSAT observation is $\sim$2.5 times its luminosity in an Einstein observation (\cite{rbarnard-C1:pfj93}, where S3 is source 58). Comparison of the 0.3--2.5 keV and 2.5--10 keV lightcurves (Fig.~\ref{rbarnard-C1_ev}) demonstrates that the low energy band accounts for nearly all photons. Near the start of observation 2, a flare in the soft flux is observed that lasts $\sim$1 ks (Fig.~\ref{rbarnard-C1_ev}).  

\cite*{rbarnard-C1:tru02} show that this object is a bright supersoft source,
having a spectrum which well is represented by a 60 eV blackbody. These
systems consist of a white dwarf suffering thermal time-scale mass transfer
from its binary companion, steadily burning the accreted gas (e.g. \cite{rbarnard-C1:kh97}). Most such SSS have been found
in the Magellanic clouds and M31 because the Galactic population is so
highly absorbed by the Galactic disk. Rather little is known about their
short-term variability, although a number are known to be transient with
turn-on times of a few days (e.g. \cite{rbarnard-C1:whi95}). We note
that the luminosity variation of such sources can be much more extreme
than the intensity variation due to the large fraction of the flux lost to
absorption.

\section{ A LMXB with a possible 3 hr orbital period}

\begin{figure}[!t]
\centering
\includegraphics[scale=0.3,angle=270]{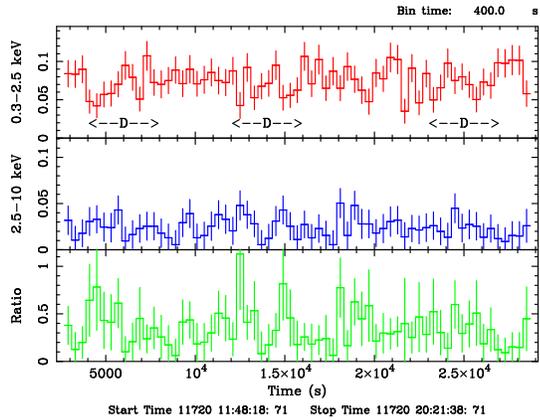}
\caption{Low (0.3--2.5 keV) and high (2.5--10 keV ) energy Obs. 1  $XMM$ lightcurves for S4, along with the ratio of the 2.5--10 keV intensity to the 0.3--2.5 keV intensity. Possible dipping events are labelled D; again, the hardness ratio is higher during dipping intervals. Note that it is quite normal for successive dips of Galactic dippers to look different}\label{rbarnard-C1_s4l}
\end{figure}

\begin{figure}[!t]
\centering
\includegraphics[scale=0.3,angle=270]{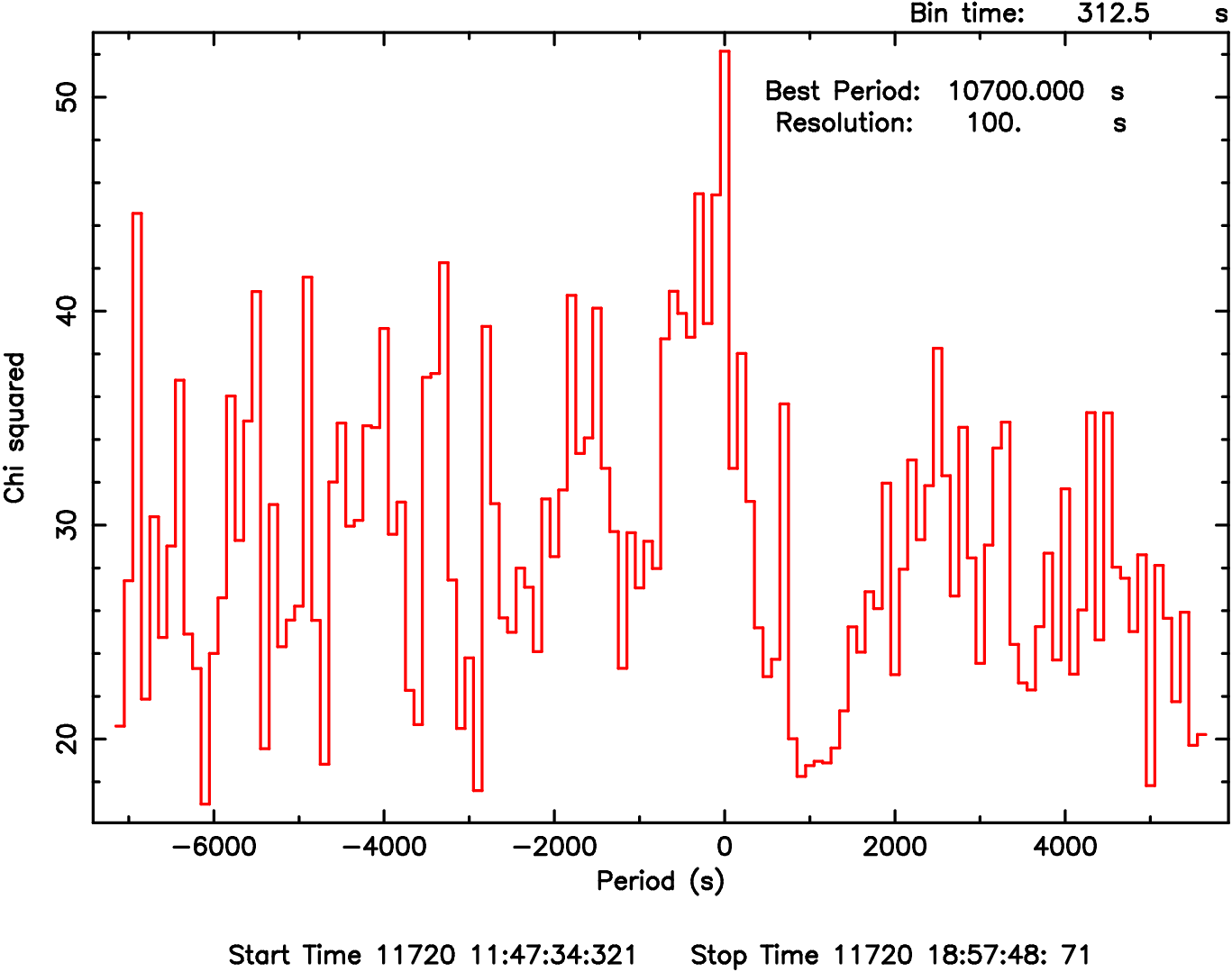}
\includegraphics[scale=0.3,angle=270]{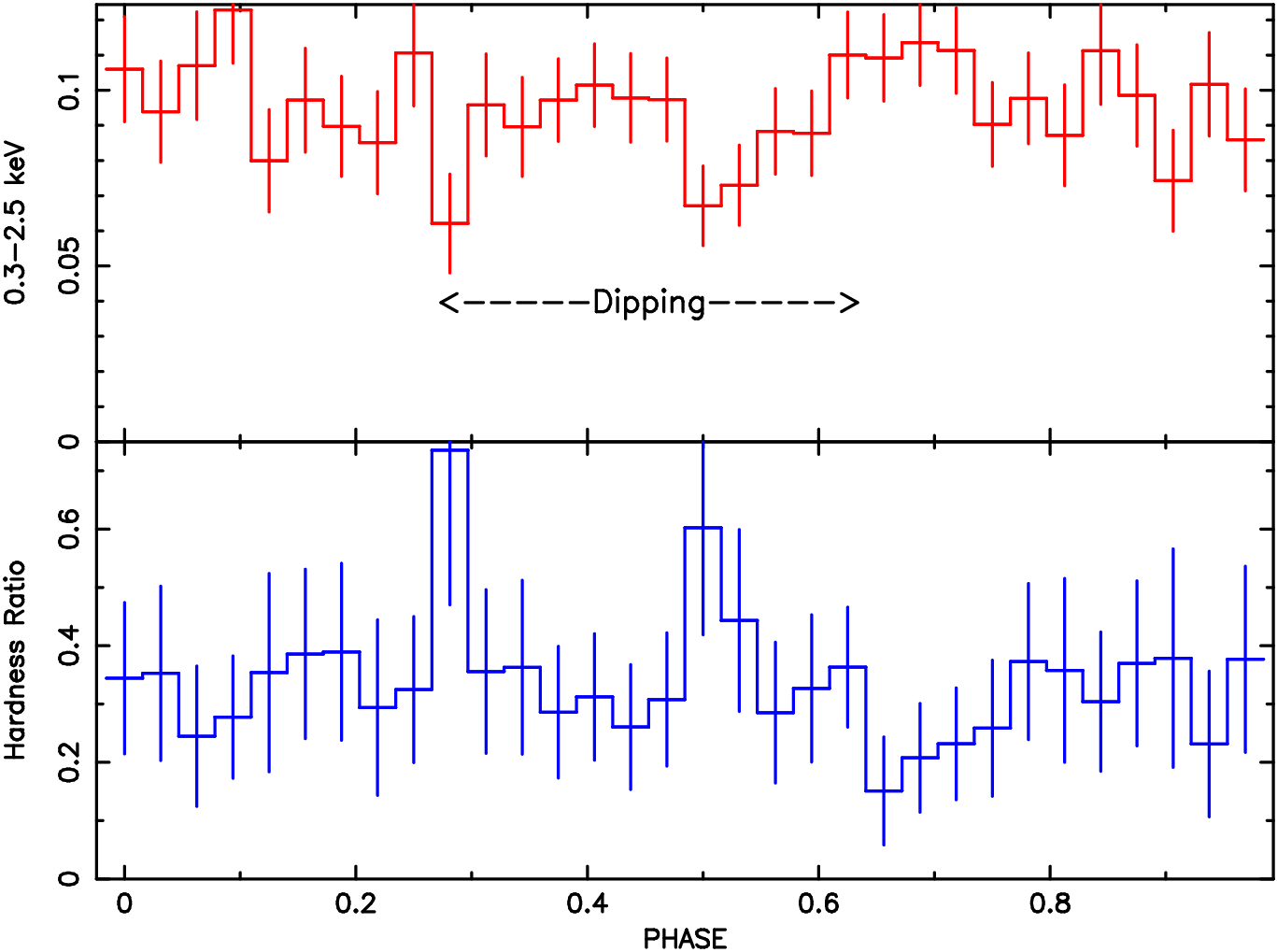}
\caption{The top plot shows the results of a period search in the 0.3--2.5 keV lightcurve of S4, using 32 phase bins and with a starting period of 10 ks. The quoted $\chi^{2}$ is for fitting a line of constant intensity to the lightcurve folded on a given period, to be compared with 31 degrees of freedom, giving a $\sim$3$\sigma$ significance for the period of 10700 s. The bottom figure shows the  lightcurve and hardness ratio (2.5--10 keV intensity divided by 0.3--2.5 keV intensity), folded on the  10.7 ks period in 32 phase bins and with an arbitrary phase; dipping is observed at phase 0.3 and $\sim$0.5--0.6, where the hardness ratio is significantly higher than in non-dip intervals}\label{rbarnard-C1_search}
\end{figure}

\object{The X\thinspace  004207.7+411813} (S4) 0.3--2.5 keV lightcurve from the 2000, June observation  exhibits several intensity dips while the high energy band  remains unchanged, indicating photo-electric absorption (Fig.~\ref{rbarnard-C1_s4l}). A period search was conducted, in which the low energy lightcurve was folded onto a range of periods, searching for the largest variability (i.e. $\chi^{2}$/d.o.f. for constant intensity). A significant preference for an orbital period of $\sim$10700 s was shown (Fig.~\ref{rbarnard-C1_search}); none of the other variable X-ray sources showed any evidence of periodicity. A folded lightcurve is presented in Fig.~\ref{rbarnard-C1_search}, and indeed there is evidence for coherent dipping, with a probability of constancy of $<$0.5\%. The modulation in the low energy lightcurve appears to be stable as each dip is consistent with the  dip in the folded curve. 

\section{ Conclusions}               

Of the 35 X-ray sources studied in the core of M31, two are dipping source candidates, one of which also exhibits a likely X-ray burst, while in the other, a possible periodicity of $\sim$3 hr has been detected with a probability of $>$99.5\%. The fraction of dipping LMXB is expected to be $\sim$20\% from inclination angle requirements, suggesting that the sample of 35 X-ray sources may contain additional dippers.   The brightest X-ray source in the field of view shows striking variability that resembles the behaviour of Galactic Z-sources, and in addition shows likely flaring, which would make it only the fifth flaring source, and the ninth Z-source. Finally, a supersoft transient source exhibits an intensity enhancement that is not yet fully understood. This early work has shown that $XMM$ is indeed capable of seeing much of the fascinating variability on offer, which may well reveal insights into the population characteristics in M31.

\end{document}